\documentstyle[12pt]{article}
                  {\def\smallskip {\vskip\smallskipamount}}
                  {\def\medskip   {\vskip\medskipamount}}
                  {\def\bigskip   {\vskip\bigskipamount}}
                  {\setbox\strutbox=\hbox{\vrule
                    height10.5pt depth4.5pt width 0pt}}
\def\middlespace {\smallskipamount=5.625pt plus1.5pt minus1.5pt
                  \medskipamount=11.25pt plus3pt minus3pt
                  \bigskipamount=22.5pt plus6pt minus6pt
                  \normalbaselineskip=22.5pt plus0pt minus0pt
                  \normallineskip=1pt
                  \normallineskiplimit=0pt
                  \jot=5.625pt
                  {\def\smallskip {\vskip\smallskipamount}}
                  {\def\medskip   {\vskip\medskipamount}}
                  {\def\bigskip   {\vskip\bigskipamount}}
                  {\setbox\strutbox=\hbox{\vrule
                    height15.75pt depth6.75pt width 0pt}}
                  \parskip 11.25pt
                  \normalbaselines}
                  {\def\smallskip {\vskip\smallskipamount}}
                  {\def\medskip   {\vskip\medskipamount}}
                  {\def\bigskip   {\vskip\bigskipamount}}
                  {\setbox\strutbox=\hbox{\vrule
                    height21.0pt depth9.0pt width 0pt}}
\title{POTENTIAL MINIMIZATION IN LEFT-RIGHT SYMMETRIC MODELS}
\author{Biswajoy Brahmachari$^1$, Manoj K. Samal$^2$, Utpal
Sarkar$^{3,4}$ \\ Theory Group, \\
Physical Research Laboratory, \\
Ahmedabad - 380009, India.}
\date{January 1994}

\begin{document}
\maketitle
\middlespace
\begin{abstract}
We study the Higgs potentials in the Left-Right symmetric model
for various choices of Higgs fields. We give emphasis to the
cases when the Higgs field $\xi=(2,2,15)$ is included to give
the correct relations of quark masses and a singlet field $\eta$
which breaks the left-right parity. As special cases we also include
$\xi^\prime=(2,2,15)$ and $\chi=(2,2,6)$ (which are interesting
in the context of the three lepton decay mode of the proton) and
field $\delta=(3,3,0)$ none of which acquire {\it vev}. We show
that the linear couplings of these fields upon minimization put fine
tuning conditions on the parameters of the model. We carry out the
minimization of these potentials explicitly. In all the cases
the relationship between the {\it vev$\;$}s of the left and right
handed triplets $v_L$ and $v_R$ are given. The phenomenological
consequences of this minimization regarding the neutrino masses
are also studied.
\end{abstract}
\rule{6cm}{.025cm}\\
1. biswajoy@prl.ernet.in\\
2. mks@prl.ernet.in\\
3. utpal@prl.ernet.in\\
4. Address during March 1994 -- February 1995 : Institut
f\"{u}r Physik, Lehrstuhl f\"{u}r Theoretische Physik III, 4600
Dortmund 50, Postfach 500500, Germany.
\newpage

\section{INTRODUCTION}
\noindent Left-Right symmetric models{\cite 1} are considered to be the most
natural extensions of the standard model. Popularly one
chooses the gauge group $G_{3221}=SU(3)_c\times SU(2)_L\times
SU(2)_R\times U(1)_{B-L}$ or $G_{224}=SU(2)_L\times SU(2)_R\times
SU(4)_c$ to describe the invariance properties of the model.
When $G_{3221}$ or $G_{224}$ admits spontaneous symmetry breaking one
recovers the standard model. Spontaneous symmetry breakdown
takes place when the Higgs fields transforming nontrivially
under the higher symmetry group but not transforming under the
lower symmetry group acquires a vacuum expectation value ({\it vev}).
If one embeds the group $G_{3221}$ or $G_{224}$ in a grand unified theory
or a partially unified theory
then LEP constraints on $sin^2\theta_w${\cite 2} can put strong bounds
on the breaking scale of the right handed $SU(2)_R$ group. On the
other hand if one considers the left-right symmetric model with
$g_L \ne g_R$ the right handed breaking scale can be
lowered{\cite 3}. In this case the model becomes interesting as a rich set of
phenomenological consequences can be directly tested in the next
generation colliders. To achieve the inequality of the couplings
a D-odd singlet Higgs field  $\eta$ is introduced which on
acquiring vev breaks the left-right parity (D-parity).

\noindent It is well known that the coupling of the Higgs field $
\phi=(2,2,1)$ under $G_{224}$ gives the masses to fermions in the left right
model. It is also known that the {\it vev} of $\phi$ alone
cannot generate the correct relationship of the quark and the
lepton masses. One has
to introduce a field $\xi=(2,2,15)$ to genetate the correct
mass relations{\cite 4}. We have included the field $\xi$ in our
Higgs potential in the presence and in the absence of D-parity
breaking. We have shown how the see-saw relationship gets
modified in the presence of the {\it vev} of the field $\xi$.

\noindent In recent past much interest is generated in the three lepton
decay mode of proton in left right symmetric model. In this
scenario one introduces the fields $\xi^\prime=(2,2,15)$ or $\chi=(2,2,6)$.
Here due to the mixing of this new fields with the field $\xi$ the
$SU(3)_c$ triplet component of $\xi$ remains light which in turn
mediate the three lepton decay of the proton with a lifetime of
$4\times 10^{31}$ years{\cite 5}. It is argued that this
decay channel can produce electron type neutrino and which may
lead to an explanation of the atmospheric neutrino problem. We
include this extra fields in special cases of our analysis. We
show that the linear couplings of these extra fields with the
other scalars present in the model can put
constraints on the parameters of the model and the {\it vev}s of
the scalar fields.

\noindent Next we consider the introduction of the scalar
$\delta=(3,3,0)$ which does not acquire {\it vev} in the
model. We show that the linear coupling of $\delta$ with other
fields in the model puts constraints on the parameters of the
model. In particular we show that in the presense of $\delta$
the scale of D-parity breaking has to be very close to the
right handed breaking scale.

\noindent In all the cases described above we perform the minimization of
the scalar potential in detail and write the
relationships among the {\it vev} s of the left handed and
right handed triplets. In most cases the {\it vev} of the left
handed triplet has to be nonzero and it contributes to the
neutrino mass\footnote{However in the presence of both $\xi$ and
$\chi$ one has to have also the presence of a D-odd singlet $\eta$
to predict an acceptable relationship between $v_L$ and $v_R$}.
The see-saw mass becomes comparable to this term.
The see-saw mass becomes compapable to this term after severely
fine tuning some parameters. This fine tuning can be avoided if
D-parity is broken. We investigate in this paper whether such
features are maintained in the presence of $\xi$.

\noindent This paper is structured as the following. In section 2 we
describe the basics of the left right model and summarize the
Higgs choices of the model. In section 3 we summarize the
results available in the literature and perform the
potential minimization in the presence of the field $\xi$. In
section 4 we introduce the fields $\xi^\prime,~\chi$, and
$\delta$. We show that the linear couplings of these fields put
strong constraints on the model parameters. In section 5 we
comment on neutrino mass matrices. In section 6 we state the
conclusions.
\section{RUDIMENTS OF LEFT RIGHT SYMMETRIC MODEL}
\noindent In this paper we are interested in the following
symmetry breaking pattern;
\begin{eqnarray}
SU(2)_L \times SU(2)_R \times SU(4)_c
& \longrightarrow \atop{M_X}&
SU(3)_c \times SU(2)_L \times SU(2)_R \times U(1)_{B-L} \nonumber\\
& \longrightarrow \atop{M_R} &  SU(2)_L \times SU(3)_c \times U(1)_Y\nonumber\\
  & \longrightarrow \atop{M_W} & SU(3)_c \times U(1)_Q
\end{eqnarray}
\noindent If $G_{224}$ is embedded in any higher symmetry group, then
also most of the analysis will not change. In this sense our
analysis is quite general. The advantage of starting with the
group $G_{224}$ instead of the group $G_{3221}$ is that, we can
discriminate between the fields which do and don't distinguish
between quarks and leptons. This is important to understand the
mass ratios of quarks and leptons.

\noindent We will also assume that $M_X=M_R$ which will imply that the
scale of breaking of SU(4) color is the same as that of the
breaking of the left right symmetry. This will not cause any
loss of generality of our analysis. To specify the model further
let us state the transformation properties of the fermions.
\begin{eqnarray}
\psi_L & = & \pmatrix{\nu_L \cr {e^-}_L}: (2,1,4)~~;~~
\psi_R  =  \pmatrix{\nu_R \cr {e^-}_R}: (2,1,4) \nonumber\\
Q_L & = & \pmatrix{u_L \cr d_L}: (2,1,4)~~;~~
Q_R  =  \pmatrix{u_R \cr d_R}: (2,1,4)
\end{eqnarray}
\noindent The scalar fields which may acquire {\it vev} are stated
below.
\begin{eqnarray}
\phi_1 & \equiv  &({2},{2},1)~~;~~
\phi_2  \equiv  \tau_2 \phi^{*}_1 \tau_2~~;~~\xi_1  \equiv  (2,2,15)~~;~~
\xi_2  \equiv  \tau_2 \xi^{*}_1 \tau_2 \nonumber\\
\Delta _L & \equiv &({3},{1},10)~~;~~
\Delta _R  \equiv ({1},{3},10)~~,~~ \eta  \equiv ({1},{1},0)   \nonumber
\end{eqnarray}

\noindent It has been shown in recent past that the LEP
constraints on $sin^2 \theta_w${\cite 2} can put strong lower bound on
the scale $M_R$. From renormalization group equations one can
show that the  right handed breaking scale has to be
greater than $10^{9}$ GeV. However one can show that when the
D-Parity is broken the right handed breaking scale can be
lowered. In that case a rich set of phenomenological predictions
can be experimentally tested in high energy colliders. Here we
consider the singlet field $\eta$ which is odd under D-Parity. It
breaks D-Parity when it acquires {\it vev}{\cite 3}.

\noindent If we consider an underlying GUT, and start with the
masses of the quarks and leptons to be the same at the
unification scale, then, in the absence of $\xi$ the low energy
mass relations of fermions are not correct. This is because the
field (2,2,1) contributes equally to
the masses of the quarks and leptons. The situation can be
corrected by the introduction of the field (2,2,15){\cite 4}. This is the
initial motivation to introduce the field $\xi$. Once it is
there it allowes new interesting baryon number violating decay
modes which we discuss below.

\noindent Recently a lot of interest has been generated in the
three lepton decay of the proton in SU(4) color gauge
theory{\cite 5}. It
can be shown that if the SU(3) triplet component of $\xi$
remains sufficiently light it can mediate the three lepton decay
mode of proton with a lifetime of $4 \times 10^{31}$ years. In
that case sufficient number of extra electron type neutrinos can
be produced in the detector which can explain atmospheric
neutrino anomaly. To keep the SU(3) triplet component of $\xi$
sufficiently light, the following mechanism  was proposed by Pati,
Salam and Sarkar. If an extra (2,2,15) or (2,2,6) Higgs field
(henceforth called $\xi^\prime$ and $\chi$) is introduced, its
SU(3) triplet component will mix with the triplet component of
$\xi$ and hence there will be a light triplet in the model. These
extra fields does not acquire {\it vev}. However the terms in
the scalar potential which are linear in these extra fields can
strongly constrain the other parameters of the model. In this
paper we introduce such extra fields which does not acquire
{\it vev} and study the terms in the scalar potential which
are linear in these extra fields. The extra fields we consider
here are,
\begin{equation}
\xi^\prime = (2,2,15)~;~
\chi = (2,2,6)~;~
\delta = (3,3,0)
\end{equation}
We shall see below that the linear term in the extra field
$\delta$ will constrain the ratio of the D-parity breaking scale
and the right handed symmetry breaking scale.
\noindent We emphasize that in different models with extra
scalars such study is necessary as it points out the extra
scalar which is not favorable by the existing phenomenology.

\section{MINIMIZATION OF POTENTIAL}
\subsection{MINIMAL CHOICE OF HIGGS SCALARS}
The general procedure we adopt here is the following. First we
write down the most general scalar potential which is allowed by
renormalizability and gauge invariance. Next we substitute the
vacuum expectation values in the potential and find out the
minimization conditions. Here let us first write down the scalar
potential with the scalar fields $\phi$, $\Delta$ and
$\eta${\cite 6},
\begin{equation}
V(\phi_1,\phi_2,{\Delta_L},{\Delta_R},{\eta}) =
V_\phi+V_\Delta+V_\eta+V_{\eta\Delta}+V_{\eta \Delta}+
V_{\eta \phi}
\end{equation}
\noindent Where the different terms in this expression are given by,
\begin{eqnarray}
&&\nonumber\\
V_\phi&=& -\sum_{i,j}\mu^2_{ij}~tr(\phi ^{\dagger}_i
\phi_j)+\sum_{i,j,k,l}\lambda_{ijkl}~tr(\phi ^{\dagger}_i
\phi_j)~tr( \phi ^{\dagger}_k \phi_l)\nonumber\\
&&+\sum_{i,j,k,l}\lambda_{ijkl}~tr(\phi ^{\dagger}_i \phi_j \phi
^{\dagger}_k \phi_l)\nonumber\\
{}~&&~\nonumber\\
V_\Delta &=&-\mu^2~(\Delta ^{\dagger}_L \Delta_L+\Delta ^{\dagger}_R
\Delta_R) +
\rho_1~[tr(\Delta ^{\dagger}_L \Delta_L)^2+tr(\Delta ^{\dagger}_R
\Delta_R)^2]\nonumber\\
&&\nonumber\\
&&+\rho_2~[tr (\Delta^{\dagger}_L \Delta_L \Delta^{\dagger}_L \Delta_L)
+tr (\Delta^{\dagger}_R \Delta_R \Delta^{\dagger}_R \Delta_R)]
+\rho_3~tr(\Delta ^{\dagger}_L \Delta_L \Delta ^{\dagger}_R
\Delta_R)\nonumber\\
{}~&&~\nonumber\\
V_\eta&=&-\mu^2_\eta~ \eta^2+\beta_1~ \eta^4 \nonumber\\
{}~&&~\nonumber\\
V_{\Delta\phi}&=&+\sum_{i,j}\alpha_{ij}~(\Delta ^{\dagger}_L \Delta_L+\Delta
^{\dagger}_R \Delta_R)~{tr(\phi ^{\dagger}_i \phi_j) }
+\sum_{i,j} \beta_{ij}~[~tr(\Delta ^{\dagger}_L\Delta_L\phi_i
\phi ^{\dagger}_j) \nonumber\\
& &+tr(\Delta ^{\dagger}_R\Delta_R\phi ^{\dagger}_i
\phi_j)]\nonumber\\
&&+\sum_{i,j} \gamma_{ij}~tr(
\Delta^{\dagger}_L\phi_i \Delta_R\phi^\dagger_j) \nonumber\\
{}~&&~\nonumber\\
V_{\eta\Delta}&=& M~\eta~(\Delta^{\dagger}_L \Delta_L-
\Delta ^{\dagger}_R \Delta_R)+\beta_2~ \eta^2~(\Delta
^{\dagger}_L \Delta_L+\Delta ^{\dagger}_R \Delta_R) \nonumber\\
{}~&&~\nonumber\\
V_{\eta\phi}&=&\sum_{i,j}\delta_{ij}~\eta^2~tr(\phi ^{\dagger}_i
\phi_j)\nonumber
\end{eqnarray}
\noindent The vacuum expectation values of the fields have the
following form:

\begin{eqnarray}
<\phi>&=& \pmatrix{k&0\cr 0& k^{\prime}}~~;~~
<\Delta_L>= \pmatrix{0&0\cr v_L&0}~~;~~<\eta> = \eta_0; \nonumber \\
<\tilde{\phi}>&=& \pmatrix{k^{\prime} &0\cr 0&k}~~;~~
<\Delta_R>= \pmatrix{0&0\cr v_R&0} \nonumber
\end{eqnarray}

\noindent The phenomenological consistency requires the
hierarchy $<\Delta_R> \; \; >> \; \; <\phi> \; \; >> \; \;
<\Delta_L>$ and also that $k^\prime << k$. Now the minimization
conditions of the potential V is found by differentiating it
with respect to the parameters $k, k^\prime, v_L, v_R$ and $\eta_0$
and separately equating them to zero. This will give us five
equations for five parameters present. Solving the equations
involving the derivatives with respect to $v_L$ and $v_R$ we get
the relation between $v_L$ and $v_R$. The details of the
derivation is presented in the appendix.
\begin{eqnarray}
v_L v_R &=& \beta k^2 \over{ [(\rho-{\rho^\prime})+ {4 M \eta_0\over
(v^2_L-v^2_R)}}] \nonumber
\end{eqnarray}
Here we have defined $\beta=2\gamma_{12}$. We get in the M=0 limit,
\begin{equation}
v_Lv_R \simeq {\beta k^2 \over [\rho - \rho^\prime] } \simeq \gamma k^2
\end{equation}
Here $\gamma$ is a function of the couplings. However when the
field $\eta$ is present, $v_L$
becomes differently related to $v_R$ in the limit of large $\eta_0$.
\begin{equation}
v_L\simeq-({\beta k^2 \over 4 M \eta_0})v_R \simeq ({\beta k^2
\over \eta^2_0}) v_R
\end{equation}
Here we see the important difference between the D-conserving
and D-breaking scenarios.

\noindent This result was discussed in details in ref.
\cite{mohch}. In the D-parity conserving case, when the $\eta$
field is absent one has to fine tune parameters to make
$\gamma$ arbitrarily snall so that the see-saw neutrino mass
can be comparable to the Majorana mass of the left-handed
neutrinos given by $v_L$. This fine tuning becomes redundant
when the field $\eta$ acquires $vev$.

\subsection{IN THE PRESENCE $\xi$=(2,2,15)}
When $\xi$ is present the most general scalar potential takes
the following form:
\begin{equation}
V(\phi_1,\phi_2,{\Delta_L},{\Delta_R},\xi_1,\xi_2,{\eta}) =
V_\phi+V_\Delta+V_\eta+V_\xi+V_{\phi \eta}+V_{\eta \Delta}+
V_{\eta \phi}+V_{\phi\xi}+V_{\Delta\xi}+V_{\eta\xi}
\end{equation}
The explicit forms of the terms involving $\xi$ are listed below,
\begin{eqnarray}
&&\nonumber\\
V_\xi&=&-\sum_{i,j}m^2_{ij}~tr(\xi ^{\dagger}_i
\xi_j)+\sum_{i,j,k,l}~n_{ijkl}~tr(\xi ^{\dagger}_i\xi_j\xi
^{\dagger}_k\xi_l)
+\sum_{i,j,k,l}p_{ijkl}~tr(\xi ^{\dagger}_i\xi_j)~
tr(\xi ^{\dagger}_k\xi_l) \nonumber\\
&&\nonumber\\
V_{\phi\xi}&=&\sum_{i,j,k,l}u_{ijkl}~tr(\phi ^{\dagger}_i \phi_j
\xi ^{\dagger}_k \xi_l)+\sum_{i,j,k,l}v_{ijkl}~tr(\phi
^{\dagger}_i \phi_j)~tr(\xi ^{\dagger}_k \xi_l) \nonumber \\
&&\nonumber\\
V_{\Delta\xi}&=&+\sum_{i,j}a_{ij}~[~tr(\Delta ^{\dagger}_L \Delta_L)+tr(\Delta
^{\dagger}_R \Delta_R)]~{tr(\xi ^{\dagger}_i \xi_j) } \nonumber\\
&& +\sum_{i,j} b_{ij}~[~tr(\Delta ^{\dagger}_L\Delta_L\xi_i
\xi ^{\dagger}_j)+tr(\Delta ^{\dagger}_R\Delta_R\xi ^{\dagger}_i
\xi_j)]\nonumber\\
&&+\sum_{i,j} c_{ij}~tr(
\Delta^{\dagger}_L\xi_i \Delta_R\xi^\dagger_j) \nonumber\\
&&\nonumber\\
V_{\eta\xi}&=&\sum_{i,j}d_{ij}~\eta^2~tr(\xi
^{\dagger}_i \xi_j)\nonumber
\end{eqnarray}
The vacuum expectation value of $\xi$ has the following form,
\begin{equation}
<\xi>=\pmatrix{{\tilde k}&0\cr 0&{\tilde k}^\prime}
\times (1,1,1,-3)
\end{equation}
Here we may briefly mention the need to introduce the field
$\xi$. The vacuum expectation value of the field
$\phi$ is given by,
\begin{equation}
<\phi>= \pmatrix{ k&0\cr 0&k^\prime}
\times (1,1,1,1)
\end{equation}
\noindent Note that in the SU(4) color space the fourth entry is
1 for the {\it vev} of $\phi$ whereas it is -3 for the {\it vev} of
$\xi$. Hence the {\it vev} of $\phi$ treats the quarks and
the leptons on the same footing, whereas the {\it vev} of $\xi$
differentiates between the quarks and the leptons. For example in
the absence of $\xi$ one gets $m^0_e=m^0_d$, $m^0_\mu=m^0_s$ and
$m^0_\tau=m^0_b$. Now including the QCD and electroweak
renormalization effects in the symmetric limit it leads to the
relation ${m_e \over m_\mu}={m_d \over m_s}$. However when the
field $\xi$ is included in the masses in the symmetric limit
they take the form $m^0_e =m^\phi_e-3m^\xi_e$ and
$m^0_d=m^\phi_d-m^\xi_d$. \\

\noindent The minimization condition is again found by taking the
derivatives of V with respect to
$k,k^\prime,v_L,v_R,\tilde{k}^2$ and ${\tilde{k^\prime}}^2 $ and
separately equating them to zero. Solving the equations
involving the derivatives of $v_L$ and $v_R$ yields in the limit
$\tilde{k}^\prime << \tilde{k}$:
\begin{eqnarray}
v_L v_R &=&[(w \tilde{k}^2+ \beta k^2)~(v^2_L-v^2_R)\over
[(\rho-\rho^\prime)(v^2_L-v^2_R)+4 M \eta_0]
\end{eqnarray}
Here we have defined $w=2c_{12}$. Let us again check the special
cases. Firstly the case without
$\xi$ can be recovered in the limit w=0, on the other hand the
case with unbroken D-parity can be restored in the limit M=0.
Which is,
\begin{equation}
v_L v_R \simeq {w \tilde{k}^2 + \beta k^2 \over
[(\rho-\rho^\prime)]}
\end{equation}
When D-parity is
broken the $v_L$ can be suppressed by $\eta_0$,
\begin{eqnarray}
v_L &=&{w {\tilde k}^2+\beta k^2\over
\eta^2_0}v_R
\end{eqnarray}
We infer that the
field $\xi$ is allowed by the potential minimization and its
introduction does not alter the general features of the see-saw
condition between $v_L$ and $v_R$.

\section{INTRODUCTION OF EXTRA FIELDS}
\subsection{INTRODUCTION OF $\xi^\prime$=(2,2,15)}
We have already mentioned that there exists interesting models
in the literature where the field $\xi^\prime$ is introduced to
induce a sufficiently large amplitude of the three lepton decay
width of the proton.  In these
models the field $\xi$ does not acquire {\it vev}. Hence after
the minimization all terms other than the ones which are linear in
$\xi^\prime$ drops out whereas the ones which are linear in
$\xi^\prime$ puts constraints on the other parameters of the
model. Usually when any new fields are introduced in any model,
which don't acquire $vev$s, it is assumed that it will not
change the minimization conditions. As a result potential
minimization with such fields were not done so far.

\noindent In this section we will first write down
what are the linear couplings of the field $\xi^\prime$.
\begin{eqnarray}
&&\nonumber\\
V_\xi^\prime&=&-\sum_{i,j}\tilde{m}^2_{ij}~tr(\xi ^{\dagger}_i
\xi^\prime_j)+\sum_{i,j,k,l}~n_{ijkl}~tr(\xi ^{\dagger}_i\xi_j\xi
^{\dagger}_k\xi^\prime_l) \nonumber\\
&&+\sum_{i,j,k,l}p_{ijkl}~tr(\xi ^{\dagger}_i\xi_j)~
tr(\xi ^{\dagger}_k\xi^\prime_l) \nonumber\\
&&\nonumber\\
&+&\sum_{i,j,k,l}u_{ijkl}~tr(\phi ^{\dagger}_i \phi_j
\xi ^{\dagger}_k \xi^\prime_l)+\sum_{i,j,k,l}v_{ijkl}~tr(\phi
^{\dagger}_i \phi_j)~tr(\xi ^{\dagger}_k \xi^\prime_l) \nonumber \\
&&\nonumber\\
&+&\sum_{i,j}\tilde{a}_{ij}~(\Delta ^{\dagger}_L \Delta_L+\Delta
^{\dagger}_R \Delta_R)~{tr(\xi ^{\dagger}_i \xi^\prime_j) } \nonumber\\
&&+\sum_{i,j} \tilde{b}_{ij}~[~tr(\Delta ^{\dagger}_L\Delta_L\xi^\prime_i
\xi ^{\dagger}_j)+tr(\Delta ^{\dagger}_R\Delta_R\xi ^{\dagger}_i
\xi^\prime_j)]\nonumber\\
&+&\sum_{i,j} \tilde{c}_{ij}~tr(
\Delta^{\dagger}_L\xi^\prime_i \Delta_R\xi^\dagger_j) \nonumber\\
&&\nonumber\\
&+&\sum_{i,j}\tilde{d}_{ij}~\eta^2~tr(\xi
^{\dagger}_i \xi^\prime_j)\nonumber
\end{eqnarray}
\noindent When this potential is minimized with respect to
$\xi^\prime$ we get a relation between the couplings and the
{\it vev} s. Obviously in this case due to large number of
couplings of the field $\xi^\prime$ (which are independent
parameters) this condition can be easily satisfied. A more
stringer and interesting situation is the case where an extra
field $\chi$ is introduced instead of $\xi^\prime$.
\subsection{INTRODUCTION OF $\chi$=(2,2,6)}
It has been pointed out by Pati{\cite 7} that the field $\chi$
is a very economical choice for the mechanism that leads to
appriciable three lepton decay of proton. The field $\chi$ is
contained in the field 54-plet of SO(10) which has to be present
for the breaking of SO(10). The terms linear in $\chi$ can be
written as:
\begin{equation}
V_\chi=P~\eta\xi\chi(\Delta_R-\Delta_L)+M~ \chi\xi(\Delta_R+\Delta_L)
\end{equation}
These terms upon minimization gives the condition
\begin{equation}
v_L={P \eta_0-M \over P \eta_0+M}v_R
\end{equation}
This means that to get $v_R>>v_L$ one has to fine tune
$P\eta_0-M <<P \eta_0+M$. This is interesting in the context of
the
three lepton decay of Proton which will be discussed elsewhere
{\cite 8}.
\subsection{INTRODUCTION OF $\delta$=(3,3,0)}
In this case we first write down the linear couplings of the
field $\delta$:
\begin{equation}
V_\delta=M_1 \delta
(\Delta_L\Delta^\dagger_R+\Delta_R\Delta^\dagger_L)+M_2 \delta
\phi \phi^\dagger+C_1 \eta \delta
(\Delta_L\Delta^\dagger_R+\Delta_R\Delta^\dagger_L) +C_2 \eta
\delta \phi^\dagger \phi
\end{equation}
These terms upon minimization gives the following conditions,
\begin{equation}
v_Lv_R=-{M_2+C_2\eta_0 \over 2 M_1+C_1 \eta_0} k^2
\end{equation}
In the limit of very large $\eta_0$ we can write,
\begin{equation}
v_L v_R \simeq k^2
\end{equation}
If we compare this relation with the see-saw relation of section
3.2 we get,
\begin{equation}
{v^2_R\over \eta^2_0}={k^2 \over w \tilde{k}^2 +\beta k^2}
\simeq O(1)
\end{equation}
Thus due to the introduction of $\delta$ the left-right parity
and the left right symmetry gets broken almost at the same scale.
\section{NEUTRINO MASS MATRIX}
The fermions acqire masses through the Yukawa terms in the
lagrangian when the Higgs fields acquire vev. The Yukawa  part
in the Lagrangian written in terms of fermionic and Higgs fields
is given by,
\begin{eqnarray}
L_{Yukawa}& =&  y_1({\bar{f_L}} f_R \phi_1)+y_2({\bar{f_L}} f_R
\phi_2) + y_3({\bar{f^c_L}} f_L \Delta_L+{\bar{f^c_R}} f_R
\Delta_R )\nonumber\\
&&+y_4({\bar{f_L}} f_R \xi_1)+ y_5({\bar{f_L}} f_R \xi_2)
\end{eqnarray}
where $y_i$ (i=1,5) are Yukawa couplings. With this notation
neutrino mass matrix written in the basis ( $\nu_L,\nu_L^c$) is
\begin{equation}
M=\pmatrix{m_{M_L}&m_D \cr m_D& m_{M_R}}\nonumber\\
\end{equation}
where $m_{M_L}$ ($m_{M_R}$) is the left (right) handed Majorana
mass term whereas $m_D$ is the Dirac mass term. These terms can
be related to the Yukawa couplings and {\it vev}s through the
following relation,
\begin{eqnarray}
m_{M_L}&=& y_3 v_L\nonumber\\
m_D &=& (y_1+y_2)(k+{k^{\prime}})+(y_4+y_5)({\tilde k}+{\tilde
{k^\prime}})\nonumber\\
m_{M_R}&=&y_3 v_R
\end{eqnarray}
Upon diagonalization of the mass matrix we obtain the mass
eigenvalues. Now let us consider the simplifying assumption that
all the Yukawa couplings are of order "h" and the {\it vev} s $k^\prime$
and ${\tilde k}^\prime$ are much smaller than the {\it vev} s $k$
and ${\tilde k}$ respectively. Under this assumption the
eigenvalues become,
\begin{eqnarray}
m_1&=&y_3 v_R \nonumber\\
m_2&=& m_{M_L}- {M^2_D \over m_{M_R}}=y_3 v_L -{h^2
(k^2+{\tilde{k}}^2) \over y_3 v_R}\nonumber
\end{eqnarray}
We substitute for $v_L$ from the see-saw condition to get in
the D-parity conserving $g_L=g_R$ case,
\begin{equation}
m_2=y_3 {(\beta k^2+w {\tilde{k}}^2)\over v_R} -{h^2
(k^2+{\tilde{k}}^2) \over y_3 v_R}
\end{equation}
We notice that the second term in the right hand side is
suppressed by the square of the Yukawa coupling. Due to this the
first term dominates. If we want to make the first term small
compared to the second we need to fine tune the parameters.
Hence one has to fine tune such that $\beta k^2+w {\tilde{k}}^2
\simeq 0$ to get acceptable value of
the the light neutrino mass. However in the presence of the
{\it vev} of $\eta$ we get,
\begin{equation}
m_2=y_3{w \tilde{k}^2+\beta k^2 \over \eta^2_0} v_R -{h^2
(k^2+{\tilde{k}}^2) \over y_3 v_R}
\end{equation}
In the limit of very large $\eta_0$ the first term drops out of
the expression and one gets rid of the fine tuning problem.
However, if the field $\delta$ (which does not
acquire any $vev$) is present, we cannot get away with the fine
tuning problem, since it is difficult to maintain $v_R <<
\eta_0$.
\section{CONCLUSIONS}
We have incorporated the scalar field $\xi$=(2,2,15) in the
scalar potential of the $SU(4)_{color}$ left-right symmetric
extension of the standard model.
This field is necessary to predict correct mass relationships of
the quarks and the leptons. After including the field $\xi$ in
the scalar potential we have carried out the minimization of
potential, and worked out the relationship between the {\it
vev}s of the left-handed and the right-handed triplets (see-saw
relationship). We have shown that the field $\xi$ is allowed by
potential minimization and its inclusion does not change the
qualitative nature of the
see-saw relationship existing in literature. Once the see-saw
relationship between the $v_L$ and
$v_R$ is known we have gone ahead to construct the neutrino mass matrix.
We have shown that even after the inclusion of the field $\xi$
one needs to fine tune the parameters in the $g_L=g_R$ case to
predict small mass for the left handed neutrino, while in the
$g_L\ne g_R$ case one naturally gets a large suppression for the left
handed neutrino mass. This happens because even after the inclusion
of the field $\xi$ the light neutrino mass gets suppressed by
the {\it vev} of the D-odd singlet $\eta$ rather than the {\it
vev} of $\Delta_R$.

\noindent If there are new scalar fields
which don't acquire any $vev$, then to check the consistency
one has to write down their linear couplings with other fields
and after minimizing the potential use the appropriate $vev$s
of the various fields. In some cases the presence of such
fields can give new interesting phenomenology. We studied some
such cases for demonstration.

\noindent In recent past it has been shown that the three lepton
decay of the proton can successfully explain the atmospheric
neutrino anomaly by producing excess of electron type neutrino
in the detector. To produce phenomenologically acceptable decay
rate in the three lepton decay mode a mechanism was suggested by
Pati, Salam and Sarkar, and later by Pati. In this mechanism one
has to include extra scalars $\xi^\prime$=(2,2,15) or
$\chi$=(2,2,6) which does not acquire {\it vev}. We have
calculated the linear couplings of such terms in the scalar
potential and shown that these terms give relations that
constrain the values of parameters and {\it vev} s of the model.
In this paper we have given these constraints. We have also
included as a special case the extra scalar $\delta$=(3,3,0) and
shown that its inclusion forces the right handed breaking scale
and the D-parity breaking scale to become almost equal.

\section*{APPENDIX}
When the spontaneous symmetry breakdown (SSB) occurs the scalar
fields acquire {\it vev}. Let us first the case when we include
only the fields $\phi$, $\Delta_L$ and $\Delta_R$. The potential
after the SSB looks like the following,
\begin{eqnarray}
V_1 &=&-\mu^2~(v^2_L+v^2_R)+ {\rho \over
4}~(v^4_L+v^4_R)+{\rho^\prime \over 4}~(v^2_Lv^2_R)
+2v_Lv_R[(\gamma_{11} \nonumber\\
&&+\gamma_{22})kk^\prime+\gamma_{12}(k^2+{k^\prime}^2)]
+(v^2_L+v^2_R)~[(\alpha_{11}+\alpha_{22}+\beta_{11})~k^2\nonumber\\
&&+(\alpha_{11}+\alpha_{22}+\beta_{22})~{k^\prime}^2
+(4\alpha_{12}+2\beta_{12})~kk^\prime]\nonumber\\
&&+{\rm terms~containing~~k~and~k^\prime~only}
\end{eqnarray}
We have defined the new parameters as
$\rho=4(\rho_1+\rho_2)$ and $\rho^\prime=2 \rho_3$.
Minimisation with respect to $v_L$ and $v_R$ yields,
\begin{equation}
v_Lv_R={2~[(\gamma_{11}
+\gamma_{22})kk^\prime+\gamma_{12}(k^2+{k^\prime}^2)] \over \rho-\rho^\prime}
\end{equation}
This expression simplifies in the limit $k^\prime << k$ to
\begin{equation}
v_Lv_R={2~\gamma_{12} \over \rho-\rho^\prime} k^2
\end{equation}
Here let us introduce the new scalar $\eta$ which has a {\it vev}
$\eta_0$. Now the scalar potential after the SSB will be,
\begin{equation}
V_2=V_1 -\mu^2_\eta~\eta^2_0+ \beta_1~ \eta_1~ \eta^4_0+M~
\eta_0~(v^2_L-v^2_R)+ \beta_2~ \eta^2_0~(v^2_L+v^2_R)+\gamma~
\eta^2_0~(k^2+{k^\prime}^2)
\end{equation}
Now the minimization with respect to $v_L$ and $v_R$ gives the
following relation in the limit $k^\prime << k$,
\begin{equation}
v_Lv_R={2~\gamma_{12} k^2\over [\rho-\rho^\prime
+{4 M \eta_0 \over (v^2_L-v^2_R)}]} ={\beta k^2\over
[\rho-\rho^\prime
+{4 M \eta_0 \over (v^2_L-v^2_R)}]}
\end{equation}
We have defined the new parameter $\beta=2 \gamma_{12}$. At this stage let us
introduce the scalar field $\xi$. This will again introduce new
terms in the scalar potential. The scalar potential after SSB now becomes,
\begin{eqnarray}
V_3&=&V_2+(v^2_L+v^2_R)~[(a_{11}+a_{22}+b_{11})~\tilde{k}^2
+(a_{11}+a_{22}+b_{22})~{\tilde{k^\prime}}^2\nonumber\\
&&+(4a_{12}+b_{12})~\tilde{k}\tilde{k^\prime}]
+ 2v_Lv_R[(c_{11}
+c_{22})\tilde{k}\tilde{k^\prime}+c_{12}(\tilde{k}^2+\tilde{k^\prime}^2)]
\nonumber\\
&&+{\rm terms~containing~~\tilde{k}~and~\tilde{k^\prime}~only}
\end{eqnarray}
Now we minimize $V_3$ with respect to $v_L$ and $v_R$. The see-saw
relation becomes,
\begin{equation}
v_Lv_R={\beta k^2+2c_{12}~(\tilde{k}^2+\tilde{k^\prime}^2)\over
[\rho-\rho^\prime +{4 M \eta_0 \over (v^2_L-v^2_R)}]}
\end{equation}
This relation in the limit $\tilde{k^\prime}<<\tilde{k}$ becomes,
\begin{equation}
v_Lv_R={\beta k^2+w~\tilde{k}^2\over [\rho-\rho^\prime
+{4 M \eta_0 \over (v^2_L-v^2_R)}]}
\end{equation}
Here we have defined $w=2c_{12}$. This is the see-saw condition
in the presence of $\xi$ .

\end{document}